\documentclass[aps,przb,twocolumn,showpacs,letterpaper,superscriptaddress]{revtex4-1}
\usepackage{amsmath}
\usepackage{amssymb}
\usepackage[utf8]{inputenc}
\usepackage{comment}
\usepackage{color}
\usepackage[usenames,dvipsnames]{xcolor}
\usepackage{soul} 
\usepackage{xr}
\usepackage{graphics}\usepackage{epsfig}\usepackage{subfigure}
\usepackage{hyperref} 
\usepackage{multirow}

\begin{document}

\title{Multi-Gap superconductivity in HgS under pressure}

\author{Pietro Maria Forcella} 
\affiliation{Dipartimento di Scienze Fisiche e Chimiche, Università degli Studi dell’Aquila, Via Vetoio 10, I-67100 L'Aquila, Italy} 
\author{Cesare Tresca} 
\email{cesare.tresca@spin.cnr.it}
\affiliation{CNR-SPIN c/o Dipartimento di Scienze Fisiche e Chimiche, Universit\`a degli Studi dell’Aquila, Via Vetoio 10, I-67100 L’Aquila, Italy}
\author{Antonio Sanna} 
\affiliation{Max-Planck-Institut f\"ur Mikrostrukturphysik, Halle (Saale), Germany}
\affiliation{Institut f\"ur Physik, Martin-Luther-Universität Halle-Wittenberg, Halle (Saale), Germany}
\author{Gianni Profeta}
\affiliation{Dipartimento di Scienze Fisiche e Chimiche, Università degli Studi dell’Aquila, Via Vetoio 10, I-67100 L'Aquila, Italy}
\affiliation{CNR-SPIN c/o Dipartimento di Scienze Fisiche e Chimiche, Universit\`a degli Studi dell’Aquila, Via Vetoio 10, I-67100 L’Aquila, Italy}

\begin{abstract}
Mercury chalcogenides is a class of materials that exhibit diverse structural phases under pressure, hosting exotic physical properties, including topological phases and chiral phonons. In particular, recent experimental results\cite{zhang2023HgSsc} on HgS reports a new superconducting phase at $\sim$21 GPa, whose origin is unknown. In this letter we theoretically investigate the pressure-induced structural phase transition in HgS and the emergence of superconductivity in the rock salt phase. Remarkably, we discover that the rock salt phase hosts a two-gap superconducting phase originating from distinct Fermi surfaces. The unusually high critical temperature of $\sim$11 K emerges naturally within this multiband scenario, highlighting the role of interband coupling beyond isotropic approximation. These results place HgS among the few systems where multiband superconductivity is observed.
\end{abstract}

\maketitle
The discovery of topological surface states in mercury monochalcogenide HgTe\cite{HgTe_b, HgTe_a} represented one of the major recent achievements in science drawing new scenarios for research in condensed matter physics. With this boost, the interest in related systems with similar topological properties have grown significantly. In particular, the topological properties of mercury monochalcogenides (HgS, HgSe, HgTe) are being increasingly investigated, with a particular focus on modifying their electronic properties and topology by applying pressure or strain\cite{zhang2023HgSsc, HgTe_QSH_Molenkamp, HgTe_bands, HgTe_crys, PhysRevB.87.075143, PJFord_1982}, thereby exploring their phase diagrams in greater depth. 

In this class of materials, mercury sulfide (HgS) is the less favorable candidate to exhibit topological features due to its structural and electronic differences with respect to the other mercury chalcogenides (HgTe and HgSe)\cite{HgTe_crys, PJFord_1982}. 

In fact, at ambient pressure, both HgTe and HgSe crystallize in a zincblende structure and are semimetals with a single Dirac point at the center of the Brillouin zone\cite{HgTe_bands,HgSe_bands}. In contrast, HgS zincblende phase results to be metastable at ambient pressure\cite{10.1063/1.332730} in favor of the cinnabar crystal structure that is the ground state of the system. The cinnabar structure is characterized by a threefold helical chains arranged in an hexagonal crystal structure, which gives rise to its chiral properties\cite{10.1038/ncomms14312,10.1021/acsnano.1c05243,Nat_chiral} (for more structural details see Supplementary Materials (SM)). In this phase, HgS is a semiconductor\cite{zhang2023HgSsc, PhysRevB.73.113201}. As a function of pressure, HgS exhibits a transition to a rock salt (RS) phase in which the helical chains are broken, the Hg-S bonds are symmetrized and the ratio between the out-of-plane and the in-plane lattice constants of the hexagonal crystal lattice becomes $\sqrt{6}$. In this new phase the system becomes metallic (see SM for details). This phase transition was deeply studied both experimentally and theoretically\cite{zhang2023HgSsc, PhysRevB.28.3330, 10.1063/1.332730, doi:10.1142/S0217984906010925, Hao_2007, PhysRevB.31.5976, PhysRevB.72.174101, PhysRevB.73.113201, SUN2006476, FanDa-Wei_2009,Perdew, shchennikov2007thermoelectric, xiao2011study} reporting a wide range of possible critical pressures, from $\sim$13-15 GPa\cite{PhysRevB.31.5976,Perdew} up to $\sim$24-29~GPa\cite{Hao_2007,shchennikov2007thermoelectric, xiao2011study,PhysRevB.73.113201}, depending from the experimental techniques and/or the theoretical details used. 

Recently, the discovery of superconductivity in the RS phase of HgS\cite{zhang2023HgSsc}, just after the phase transition from the cinnabar phase, opens new scientific scenarios in the high-pressure physics of this compound. The measured critical temperature of $\sim$11~K at 25.4~GPa\cite{zhang2023HgSsc} results to be the highest critical temperatures in RS-type metal sulfide superconductors\cite{21092007, 10.1007/s10948-014-2632-y, GUPTA20201353714}. 

The discovery of such high critical temperature in Hg-based compounds is puzzling and partially unexpected\cite{CLAESON19661081, Hg-stuff-1, Hg-stuff-2},  still waiting for a complete understanding of its origin.
With this aim,  we studied the pressure driven superconducting phase in HgS by first-principles Density Functional Theory (DFT)~\cite{QEcode,QE-2017,QE-2020} and Super-Conducting Density Functional Theory (SCDFT)\cite{PhysRevB.72.024545, PhysRevB.72.024546} approach (all the computational details can be found in SM), revealing that the superconducting phase of HgS is very peculiar with the critical temperature boosted  by two distinct $s$-wave superconducting gaps.

Two-gap superconductors (of which MgB$_2$ is the most notable example) are rather rare presenting two superconducting energy gaps in the excitation spectra, related to different pairing mechanisms on different bands\cite{MgB2-origin-Nature}. When this scenario occurs it often leads to an increase in the superconducting critical temperature\cite{MAZIN200349, 10.1007/s10948-005-0052-8} resulting in interesting physical properties\cite{PhysRevLett.119.097002}.

As anticipated, the pressure phase diagram of HgS hinders subtle aspects which contributed to discrepancies in both experimental and theoretical studies in the determination of the cinnabar to rock salt transition pressure.
In particular, the calculated equation of state under pressure using the PBE exchange-correlation functional reported a transition pressure of 15.8 GPa\cite{Perdew}, while, using the meta-GGA SCAN exchange-correlation functional, the transition pressure increases up to 21.9 GPa\cite{Perdew}, much closer to the recent experimental observations by H. Zhang \textit{et al.}\cite{zhang2023HgSsc} ($\sim$21 GPa). 
\begin{figure}[h!]
\centering
\includegraphics[width=0.99\linewidth]{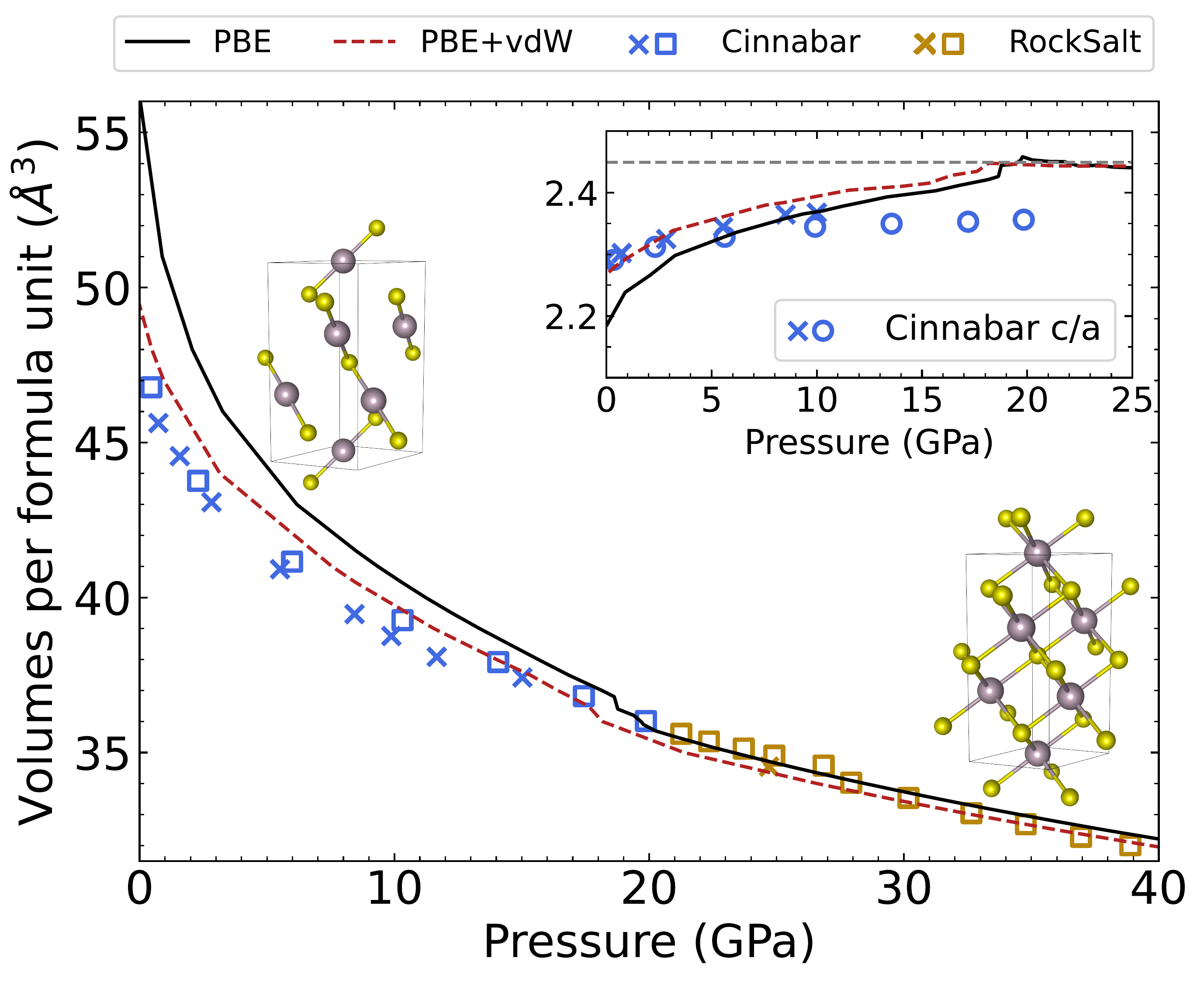}
\caption{Volume as a function of pressure. The solid black line represents the calculations performed using PBE exchange-correlation functional and the dashed red line the calculations performed using PBE exchange-correlation functional with vdW corrections. Symbols indicate the experimental values: $\square$ from Ref.\cite{zhang2023HgSsc}, $\times$ from Ref.\cite{PhysRevB.28.3330}. Blue and gold symbols refer to the cinnabar and RS phase, respectively, as assigned by the authors (please note that the $\times$ point at highest pressure has not been unequivocally attributed to the RS phase by the authors). In the inset are presented the  $c/a$ ratio as a function of pressure, computed with DFT (solid lines) and from the experimental values (blue $\circ$ and $\times$ ). The dashed gray line represents indicates $c/a$=$\sqrt6$.
}
\label{V_P}
\end{figure}
Although this evidence points to the need of meta-GGA functional to describe the high-pressure physics of HgS, we decided to reconsider the problem using different functionals and tight computational parameters (see SM for details).
Considering the helical structure of the cinnabar phase, we corrected the PBE functional  including van der Waals (vdW) contribution to account for interactions between the chains.

We report the calculated HgS volume as a function of pressure, see Fig.\ref{V_P}, using  PBE and PBE+vdW corrections, compared with the experimental results\cite{PhysRevB.28.3330,zhang2023HgSsc} (blue and gold symbols for the cinnabar and RS phases, respectively). Theoretical transition pressure has been estimated as the pressure at which the $c/a$ ratio reaches $\sqrt{6}$ (see the inset of Fig.~\ref{V_P}). This marks the point where the helical structure of the cinnabar phase is fully suppressed and the system evolves continuously into the more symmetric RS phase. In fact, the cinnabar phase can be viewed as a RS structure with the $c$ axis compressed along  the $[1 1 1]$ direction of the primitive RS crystal lattice (details on the crystal lattice and the transition from the cinnabar to the RS phase can be found in SM). 

As expected, the PBE functional fails in predicting the equilibrium volume and the equation of state for the semiconducting cinnabar phase, while the metallic RS phase is properly described. Van der Waals corrections strongly improves the theoretical results which are now in excellent agreement with the experimentally measured values (Fig.~\ref{V_P} dashed-red line). 

At variance with the calculations reported by C. Shahi \textit{et al.}\cite{Perdew}, we find that the transition pressure obtained with PBE functional is in good agreement with what predicted using the SCAN functional and with the experimental data from H. Zhang \textit{et al.}\cite{zhang2023HgSsc} (21 GPa).

We want to underline the importance of computational details in the determination of the right transition pressure: due to the low energy gap of the cinnabar phase and its closure increasing the pressure,  high electronic temperature results in a lower transition pressure. Properly converging this value, both functionals agree in predicting the critical transition pressure (see SM). 

The structural phase transition from cinnabar to rock salt is accompanied by a semiconducting to metal transition\cite{zhang2023HgSsc}, which is confirmed by our band structure calculations (see SM). Experiments also report an indirect to direct-gap transition highlighted by discontinuous slope of the optical gap as a function of the pressure\cite{zhang2023HgSsc} which is also well predicted by calculations of the band gap evolution as a function of pressure (see SM).

\begin{figure}[h!]
\centering
\includegraphics[width=0.99\linewidth]{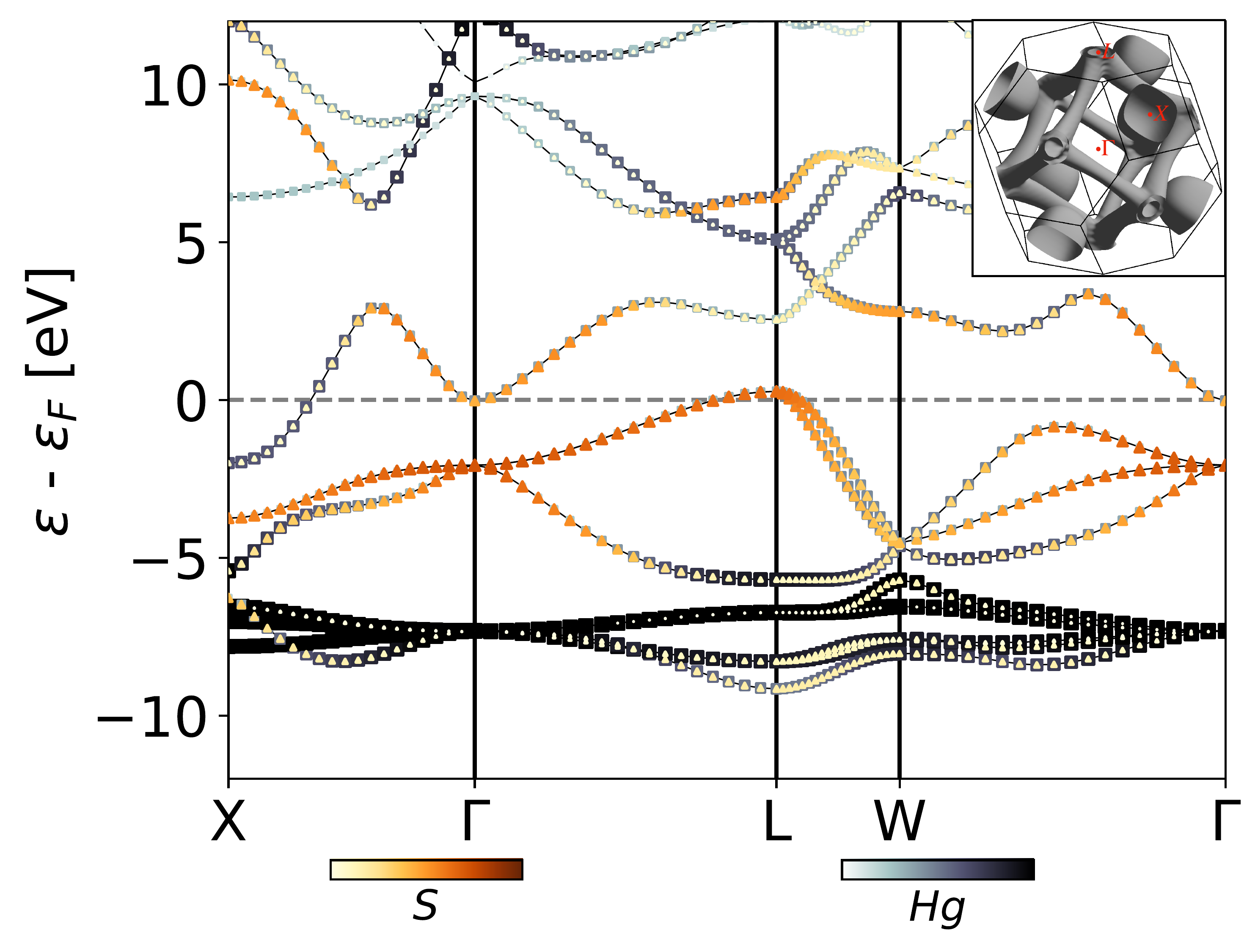}
\caption{Electronic band structure of  HgS in the RS-phase at 26~GPa calculated at PBE level. The red triangles represent the weight of the S-derived states, while the black squares of the Hg-derived states. Energies are referred to the Fermi level. The inset displays the Fermi surface of the system.}
\label{HgSHSE}
\end{figure}

When the system enters the RS phase, the electronic band structure, reported in Fig.~\ref{HgSHSE}, shows a metallic behavior with $s$-Hg electron-like band and $s/p$-S hole-like derived bands cutting the Fermi level ($\varepsilon_F$) near the $X$ and $L$ point of the Brillouin zone (BZ), respectively (see SM for further details of the band structure).
Interestingly, the hole-like and electron-like bands give rise to distinct sheets of the Fermi surface (FS), characterized by nearly spherical pockets at the $X$ point and S-derived tubular-like sheets connecting the equivalent $L$ points of BZ (see inset in Fig.~\ref{HgSHSE}). These sheets do not cross and remain separate throughout the entire BZ.
Furthermore, we point the presence of an incipient band very close to the Fermi level at $\Gamma$ which, however, has a negligible density of states.

\begin{figure}[h!]
\centering
\includegraphics[width=0.99\linewidth]{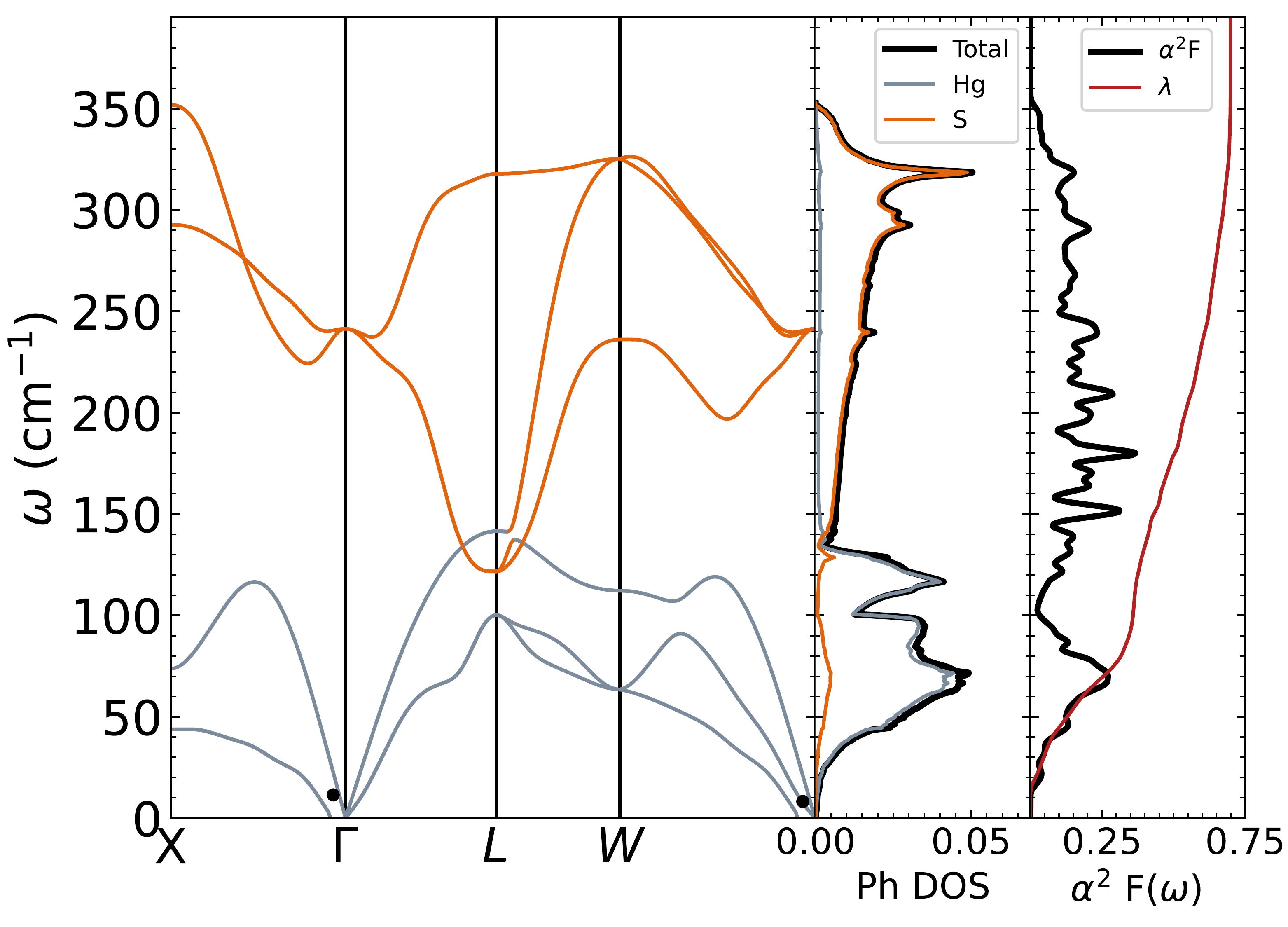}
\caption{The harmonic phonon dispersion of the RS phase at 26 GPa, calculated using the PBE exchange-correlation functional is reported in the left panel. Grey lines indicate Hg modes and orange ones the S modes. The two black dots represent the calculated phonon frequencies at near-$\Gamma$ $q$-point without interpolation (see text). In the central panel the projected phonon density of states is shown. In the right panel the Eliashberg function ($\alpha^{2}F$) is reported in black. The solid red line is the electron-phonon coupling parameter as a function of frequency.}
\label{Bands_25}
\end{figure}

We explore the possible electron-phonon origin of the observed superconducting phase studying the dynamical properties of HgS at 26~GPa. The phonon dispersion and density of states are presented in Fig.~\ref{Bands_25}.
The phonon modes show a separation between the low frequency modes, below 150~cm$^{-1}$, involving essentially Hg atoms and the optical S-derived modes, which slightly overlap around 120--150~cm$^{-1}$ energy range, as highlighted by the pseudo-gap in the phonon density of states.
Interestingly, Hg-derived modes resembles that observed in pure Hg metal at ambient conditions, extending up to $\sim$130~cm$^{-1}$\cite{TrescaHg}. Nevertheless, the characteristic strongly coupled low-energy branch which drives the superconducting phase in pure Hg\cite{TrescaHg} is now absent. 

The system presents many softenings: one in the Hg-acoustic phonon branches around the $X$ and $W$ points of the BZ, as well as along the $\Gamma$--$L$ path; and a more evident anomaly in the S-optical mode at the $L$ point, which causes the overlap between S- and Hg-modes. This overlap has a non-trivial aspect related to the character inversion between branches, as discussed in SM. 
The slight instability observed close to $\Gamma$ (along the $\Gamma$--X and W--$\Gamma$ directions) originates from numerical aspects related to the fast Fourier transform interpolation to calculate the dispersion. In fact, first-principles calculations of the phonon frequency at the corresponding $\mathbf{q}$-point (black points in Fig.~\ref{Bands_25}) results in $\omega = \text{11}$ cm$^{-1}$ indicating that the system is indeed dynamically stable (this mode, however, has a negligible electron-phonon contribution).
Notably, the RS phase remains dynamically stable even at lower pressure, below the thermodynamic transition pressure (19.6~GPa) up to 15.4~GPa where the system becomes dynamically unstable (see SM). This evidence suggests that the RS phase could persist as a metastable configuration in the $\sim$ 15–20~GPa range opening to the possibility of enhancing electron-phonon coupling in this intermediate pressure regime.

The Eliashberg spectral function, shown in Fig.~\ref{Bands_25}, yields an electron-phonon coupling constant of $\lambda = 0.70$. A significant contribution to this value arises from the softening of the lowest acoustic mode along the high-symmetry $\Gamma$--$L$ direction, around 70~cm$^{-1}$. Both Hg- and S-derived phonon modes contribute roughly equally to the total coupling strength.
The $\omega_{log}$ parameter is 12.41 meV and the estimation of the superconducting critical temperature with the Allen-Dynes formula \cite{AllenDynes} gives $T_c =$ 5~K (with $\mu^*=0.1$), a value that is notably lower than the 11~K experimentally observed\cite{zhang2023HgSsc}. 
This discrepancy between theoretical and experimental T$_c$'s, well beyond the relative error of the Allen-Dynes formula\cite{AllenDynes}, calls for a deeper reconsideration of the problem. 

Considering the peculiar Fermi surface of the system formed by distinct sheets with different orbital character, we explored the possibility of a multigap superconducting phase in HgS, in analogy with what observed in MgB$_2$ that exhibits comparable $\lambda$\cite{FLORIS200745}. In fact, multigap superconductivity can arises in materials where multiple distinct bands contribute to the formation of the superconducting state. Each band, associated with a different Fermi surface sheet, can exhibit its own superconducting gap, which may vary in magnitude and phase depending on the interactions within and between the bands, and in general can enhance the superconducting critical temperature with respect to the isotropic superconducting phase.

\begin{figure}[h!]
\centering
\includegraphics[width=0.99\linewidth]{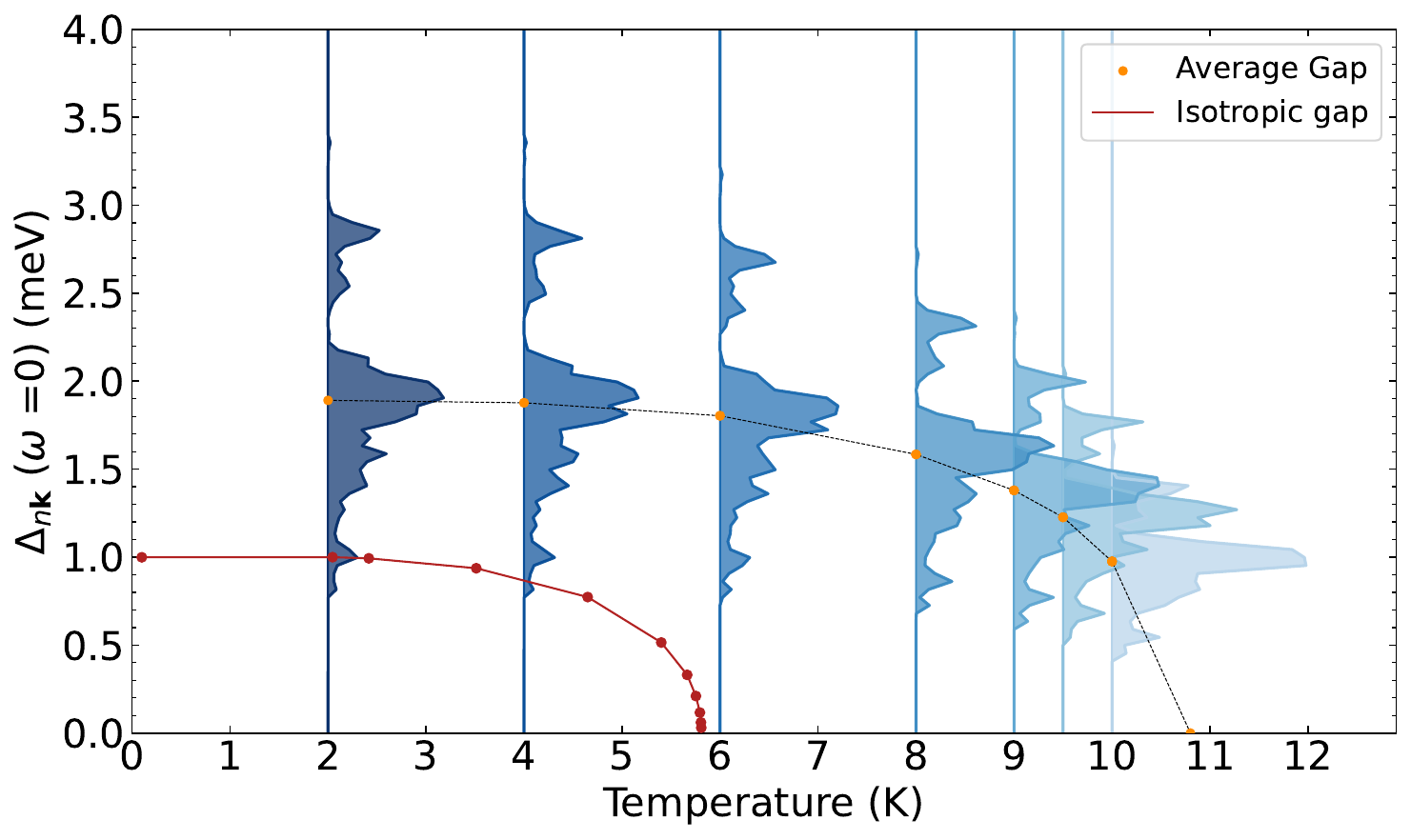}
\caption{Solution of the SCDFT anisotropic equation for the $\mathbf{k}$-resolved superconducting gap as a function of temperature.  A histogram  shows the distribution of the superconducting gap as a function of the energy for each temperature. Orange points represent the "averaged" gap at each temperature and the dashed black-line connecting them is a guide for the eyes. 
In red, the isotropic solution.}
\label{ME_anISO}
\end{figure}

\begin{figure}[h!]
\centering
\includegraphics[width=0.9\linewidth]{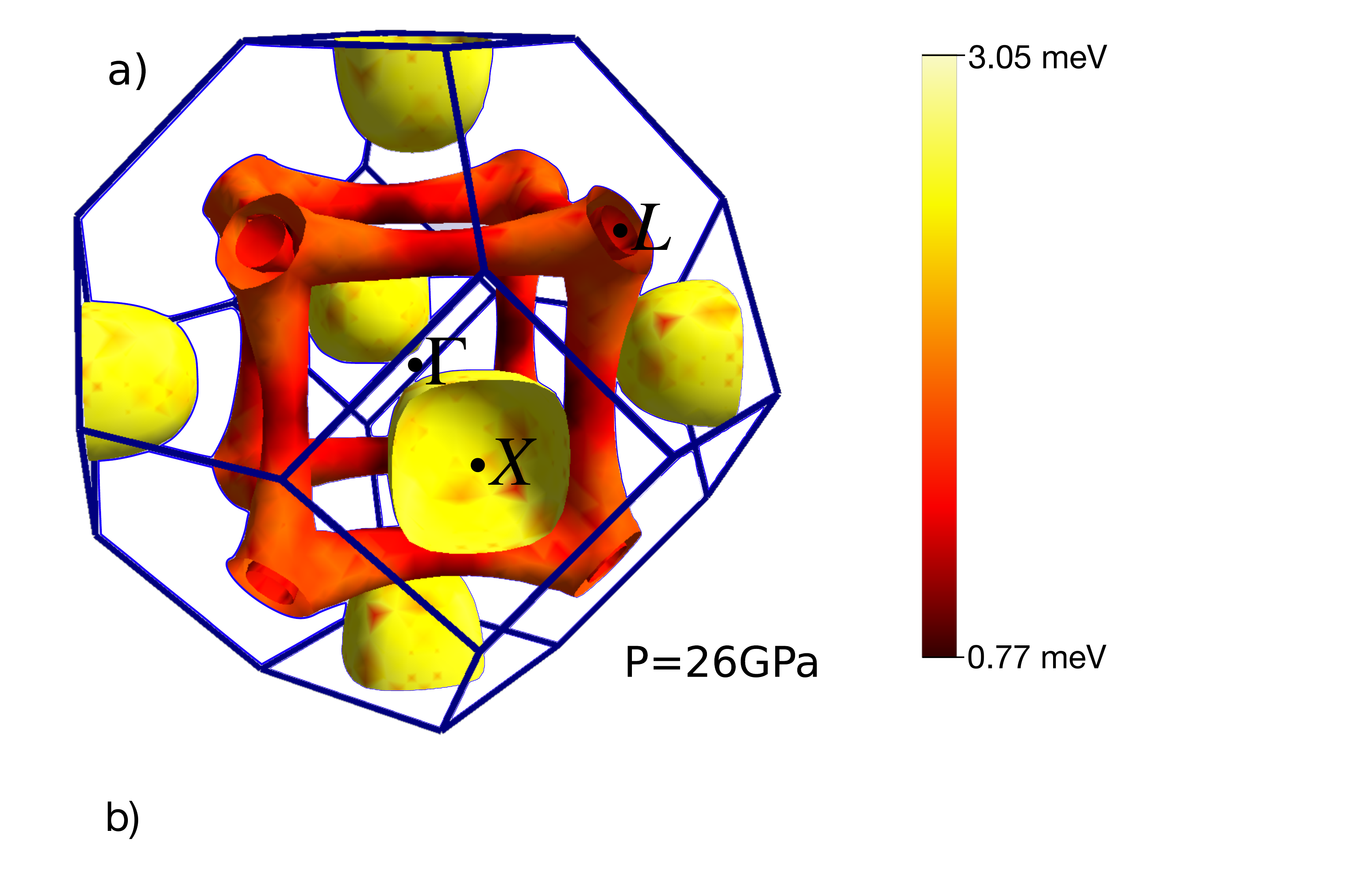}
\includegraphics[width=0.9\linewidth]{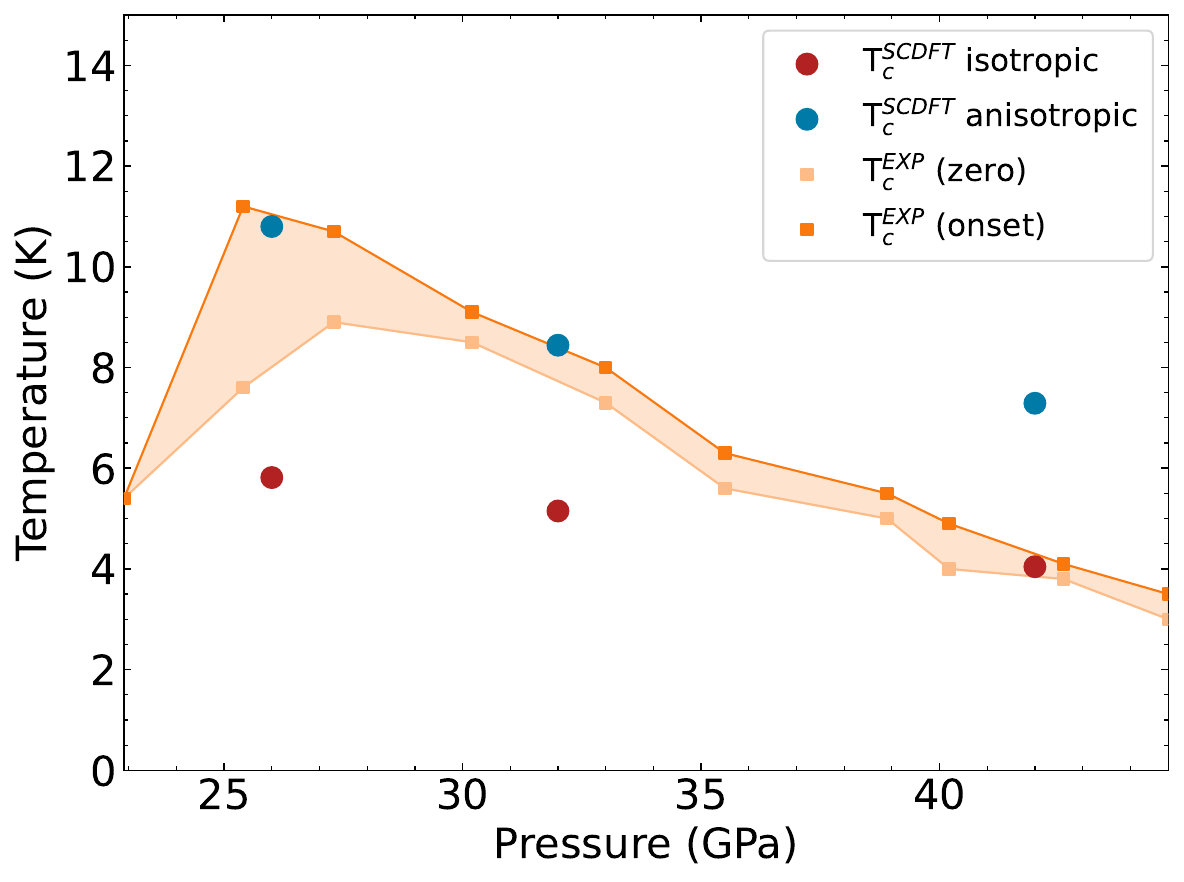}
\caption{(a) Fermi surface of RS-HgS system. The colorbar represents the magnitude of the SCDFT gap. (b) The superconducting critical temperature as a function of pressure estimated from the SCDFT isotropic (red) and anisotropic (blue) calculations. The shaded orange region highlights the experimental range between the T$_c$ onset and zero resistivity temperature\cite{zhang2023HgSsc}.}
\label{SCDFT}
\end{figure}

In order to explore this solution we studied the superconducting phase with the first-principles Superconducting Density Functional Theory (SCDFT) \cite{PhysRevB.72.024545, PhysRevB.72.024546} method, treating both attractive electron-phonon and repulsive electron-electron interactions on equal footing. 
In Fig. \ref{ME_anISO}, we report the isotropic solution of the SCDFT gap equation (red line) predicting T$_c$=5.8~K, in line with the Allen-Dynes estimation. 
In the same Figure (blue-gradient histograms) we present the results of the anisotropic solution of the SCDFT gap equation as a function of the temperature.
The multigap nature of the superconducting state in RS-HgS is clearly visible: at $T = 2$~K, two distinct peaks appear at 1.7~meV and approximately 2.7~meV. 
The two superconducting gaps are associated with the two different (disconnected) FSs sheets (see Fig. \ref{SCDFT}(a)). The larger gap is derived by the electron-like band at the $X$-point, while the other one is due to the hole-like FS's sheet. Additionally, a tiny isolated peak is observed around 3.3~meV, originating from the incipient electron pocket at the $\Gamma$ point of the Fermi surface, whose contribution for the superconducting critical temperature is negligible due to the low density of states.
Notably, the superconducting critical temperature, defined as the point where both the superconducting gaps reach zero, yields a T$_c$=10.8~K. 
The predicted critical temperature is now in excellent agreement with experimental value showing that, as for MgB$_2$, the multigap physics is responsible for the sizable increase of the superconducting critical temperature.

We extended the calculation to the whole pressure range in which the superconducting phase was measured\cite{zhang2023HgSsc}, reporting the calculated T$_c$ compared to the experimental measurements in Fig.\ref{SCDFT}(b) (we report a rough estimation of both the on-set temperature 
and the "zero resistivity" temperature as graphically extracted from Fig. 3(b) of Ref.\cite{zhang2023HgSsc}). 
The experimental superconducting critical temperature is maximum at pressures close to the structural phase transition, and then decreases almost linearly.
Our calculations confirm this trend with a good agreement up to about 33~GPa, Fig. \ref{SCDFT}(b).
However, as we can see, the agreement between theoretical predictions and experimental results at higher pressure worsens. Interestingly, for the high pressure region the isotropic solution seems to agree better with the measured T$_c$'s. At the moment, we do not have a physical explanation for this behavior, but we underline how the calculation of the critical temperature was performed using the state-of-the-art methodology and,  if  experimentally confirmed, it  will represent a challenge for theoreticians. 
However, given the sensitivity of multigap superconductors to interband scattering caused by impurities \cite{PhysRevB.55.15146, PhysRevB.50.3266, PhysRevB.50.15967, PhysRevB.53.12462}, a plausible hypothesis could be related to the experimental difficulties in maintaining sample purity (defect free) at elevated pressures. The presence of structural defects  could increase the interband coupling averaging the gap over the different bands and thus decreasing the multiband character, eventually lowering the critical temperature.

In conclusion, in this study we have provided a comprehensive theoretical investigation into the pressure-dependent phase diagram and superconducting properties of mercury sulfide. 
Our theoretical work predicts the HgS transition from its ambient pressure cinnabar phase to the metallic RS structure in line with experimental observations. Furthermore, our exploration of the phonon dynamics revealed key insights into the modes driving the structural transition, with implications for superconductivity. We reveal the multigap superconducting nature of the RS phase  with predicted critical temperatures resembling the experimental observations in a wide range of pressures. However, at high pressures ($\gtrsim$40~GPa), the multigap solution overestimates T$_c$ and the experiments seems to be explained by the isotropic solution, probably due to the emergence of pressure-induced defects. We call for further experiments to better clarify these aspects. 

\section*{Acknowledgments}
C. T. acknowledges financial support under the National Recovery and Resilience Plan (NRRP), Mission 4, Component 2, Investment 1.1, 
funded by the European Union – NextGenerationEU– Project Title “DARk-mattEr DEVIces for Low energy detection - DAREDEVIL” – CUP D53D23002960001 - Grant Assignment Decree No. 104  adopted on 02-02-2022 by the Italian Ministry of Ministry of University and Research (MUR) and Università degli Studi di Perugia and MUR for support within the project Vitality.\\
C. T.  acknowledges financial support under the National Recovery and Resilience Plan (NRRP), Mission 4, Component 2, Investment 1.1, funded by the European Union – NextGenerationEU– Project Title "Symmetry-broken HEterostructurEs for Photovoltaic applications - SHEEP" – CUP B53D23028580001 - Grant Assignment Decree No. 1409  adopted on 14-09-2022 by the Italian Ministry of Ministry of University and Research (MUR).\\
C. T., P. M. F. and G. P. acknowledge support from Laboratori Nazionali del Gran Sasso for computational resources.\\
G. P. acknowledges financial support by the European Union – NextGenerationEU, Project code PE0000021 - CUP B53C22004060006 - “SUPERMOL”, “Network 4 Energy Sustainable Transition – NEST”.\\
Research at University of L'Aquila and SPIN-CNR has been funded by the European Union - NextGenerationEU under the Italian Ministry of University and Research (MUR) National Innovation Ecosystem grant ECS00000041 - VITALITY.\\
\\

\bibliographystyle{unsrt}
\bibliography{bib}
\end{document}